\documentclass[prl,amssymb,twocolumn,tightenlines,showpacs]{revtex4-1}

\usepackage{graphicx}
\usepackage{amsmath}
\usepackage{verbatim}
\usepackage{subfigure}

\begin{document}

\title{The Effect of Decoherence on the Contextual and Nonlocal Properties of a Biphoton}

\author{A. Shaham}
\affiliation{Racah Institute of Physics, Hebrew University of
Jerusalem, Jerusalem 91904, Israel}
\author{H. S. Eisenberg}
\affiliation{Racah Institute of Physics, Hebrew University of
Jerusalem, Jerusalem 91904, Israel}


\begin{abstract}
Quantum contextuality is a nonintuitive property of quantum
mechanics, that distinguishes it from any classical theory. A
complementary quantum property is quantum nonlocality, which is an
essential resource for many quantum information tasks. Here we
experimentally study the contextual and nonlocal properties of
polarization biphotons. First, we investigate the ability of the
biphotons to exhibit contextuality by testing the violation of the
KCBS inequality. In order to do so, we used the original protocol
suggested in the KCBS paper, and adjusted it to the real scenario,
where some of the biphotons are distinguishable. Second, we
transmitted the biphotons through different unital channels with
controlled amount of noise. We measured the decohered output states,
and demonstrated that the ability to exhibit quantum contextuality
using the KCBS inequality is more fragile to noise than the ability
to exhibit nonlocality.
\end{abstract}

\maketitle

Quantum theory has a unique probabilistic nature, different from any
classical theory. It is manifested by a surprising behavior that
doesn't have any classical analogue: while measurements performed on
classical systems have predefined outcomes, compatible with a
non-contextual hidden variable assumption, quantum systems are
contextual -- the measurement outcome of a quantum system does
depend on the measurement arrangement and the choice of observables.
This contradiction between quantum mechanics predictions and
non-contextual hidden variable models, is manifested in the
Kochen-Specker theorem \cite{Kochen_Specker_1967}, and can be tested
using quantum systems with a Hilbert space dimension of $d\geq3$
\cite{Kochen_Specker_1967,Bell_1966}. A particular example for this
conflict is the famous Einstein-Podolsky-Rosen (EPR) paradox
\cite{EPR_1935}: when measurements are performed on two spatially
separated quantum subsystems, one may observe nonlocal correlations
which can not be reproduced by any local hidden variable model.

In 1964, John Bell proposed an experimental test in which one can
refute the existence of local hidden variables, by violating an
inequality, which is satisfied by classical theories
\cite{Bell_1964}. Later on, the CHSH inequality - a simpler version
of Bell inequality, which is suitable for two-qubit entangled
states, was derived \cite{CHSH_1969}. This inequality has been
tested experimentally, favoring the nonlocal predictions of quantum
mechanics \cite{Aspect_1982}. With the growing interest in quantum
information, it was discovered that states that can violate Bell
inequalities are important for many quantum processing techniques
such as quantum teleportation and quantum cryptography
\cite{Nielsen}. The Horodecki criterion for the optimal observables
for the CHSH operator should be used in order to evaluate whether a
given state can violate the CHSH inequality and thus, serve as a
resource for such tasks \cite{Horodecki_1995}.

In his response to the EPR paper, Bohr pointed out that EPR-like
peculiarities can be observed without a spatial separation between
the system components \cite{Bohr_1935}. This was rigorously proved
by Kochen and Specker \cite{Kochen_Specker_1967}. Later on,
proposals for experiments (i.e., the manifestation of non-contextual
inequalities) that are aimed to distinguish between quantum
mechanical predictions and non-contextual realism have been
suggested \cite{Cabello_1998,Simon_2000,Cabello_2008}. A special
focus was on the quantum features of three-level systems (qutrits),
which are the simplest systems that can exhibit contextual
correlations, but not nonlocal ones. Inconsistency between quantum
predictions for measurements performed on qutrits and classical
assumptions has already been described by Wright \cite{Wright_1978}.
More recently, a contextual inequality which is suitable for qutrits
was derived by Klyachko, Can, Binicio\u{g}lu, and Shumovsky (KCBS)
\cite{KCBS_2008}. This inequality is state dependent, and was proved
to be the simplest contextual inequality. In its geometrical form,
the KCBS inequality resembles Wright inconsistency, giving it a
contextual interpretation. So far, contextual inequalities were
tested experimentally using several realizations of quantum systems
including photonic systems \cite{Huang_2003,Liu_2009,Amsalem_2009},
neutrons \cite{Hasegawa_2003}, trapped ions \cite{Kirchmair_2009},
and nuclear spins \cite{Moussa_2010}. Specifically, a violation of
the KCBS inequality using indivisible photonic qutrits, encoded in
the spatial and polarization degrees of freedom, was reported in
Ref. \citenum{Lapkiewicz_2011}. A KCBS violation was also used to
certify the randomness of a random number generator, based on the
measurements of a single trapped ion qutrit \cite{Um_2013}.
Recently, Ahrens \textit{et al.} reported on two experiments where
in the first one, Wright inconsistency was demonstrated using
multi-mode single-photon qutrits \cite{Ahrens_2013}. In the second
one, the same qutrit source was used to achieve a KCBS violation
with the minimal required number of projections. A violation of the
geometrical form of the KCBS inequality has also been reported by
Kong \textit{et al.} who implemented the qutrits using NV centers
\cite{Kong_2012}.

In this work, we investigated the contextual and nonlocal properties
of biphotons - pairs of photons that occupy the same spatio-temporal
mode \cite{Burlakov_1999}. We encoded the information in the
polarization degree of freedom. Due to the symmetry of such state,
it is confined to the two-qubit triplet subspace. One option to span
this subspace is with three out of the four maximally entangled Bell
states, when the singlet is omitted. Thus, the same system can
exhibit either contextual properties, or nonlocal properties when
the two photons are divided. Recently, Soeda \textit{et al.}
suggested a simple hierarchy between contextuality and nonlocality
\cite{Soeda_2013}. They showed that contextuality with respect to
the KCBS inequality implies nonlocality, when the two-photons are
separated, while the opposite is not true.

In the first part of this letter, we use the projection protocol
suggested by Klyachko \textit{et al.} \cite{KCBS_2008} to test the
biphoton's ability to violate the KCBS inequality in its geometric
representation. We find that distinguishability imperfections of the
biphoton \cite{Adamson_2007} dramatically affect the measured KCBS
value. Thus, it is important to quantify these imperfections in
order to faithfully demonstrate contextuality. In the second part,
we explore the effects of decoherence on the biphoton ability to
exhibit different aspects of quantumness. The biphotons have passed
through three major examples of controlled unital channels
\cite{Shaham,Shaham_iso_depo}, and the biphoton output state was
characterized by a quantum state tomography procedure. We study the
nonlocal and the contextual properties of partially polarized
biphotons, and their relative hierarchy.

A general qutrit wave function of a biphoton state $|\psi\rangle$
can be written as
\begin{equation}\label{GeneralPsi}
|\psi\rangle=\alpha_0|2,0\rangle+\alpha_1|1,1\rangle+\alpha_2|0,2\rangle\,,
\end{equation}
where $|n_h,n_v\rangle$ represents a two-photon state that is
composed of $n_h$ horizontally polarized photons and $n_v$
vertically polarized photons. A qutrit state is defined as neutrally
polarized if its spin projection onto the Z axis is zero. A common
qutrit basis of neutrally polarized states is composed of the
following states \cite{Burlakov_1999}:
$\{|\psi_{hv}\rangle\equiv|1,1\rangle,|\psi_{pm}\rangle\equiv(|2,0\rangle-|0,2\rangle)/\sqrt{2},|\psi_{rl}\rangle\equiv(|2,0\rangle+|0,2\rangle)/\sqrt{2}\}$.

Consider the geometric form of the KCBS inequality \cite{KCBS_2008}
\begin{equation}\label{KCBS_eq}
K=\sum_{k=1}^5|\langle{l_k}|\psi\rangle|^2\leq2\,,
\end{equation}
where $|l_k\rangle$ are five qutrit states that satisfy
$|l_k\rangle\perp|l_{k+1}\rangle$ ($k+1$ modulo 5), and
$|\psi\rangle$ is the qutrit state of interest. Since this
inequality is state-dependent, different $|\psi\rangle$ states
require different sets of $|l_k\rangle$ to achieve maximal $K$
values. There are some qutrit states, like the $|2,0\rangle$ state,
that can not exhibit violation for any chosen set of $|l_k\rangle$.
It can be shown that by choosing $|\psi\rangle$ to be neutrally
polarized, and by choosing a quintuplet of $|l_k\rangle$ such that
$|\langle{l_i}|\psi\rangle|=|\langle{l_j}|\psi\rangle|$ for any $i$
and $j$, a maximal violation of Eq. (\ref{KCBS_eq}) is obtained,
where the KCBS value is $K=\sqrt5\approx2.24>2$.

Given a qutrit state, one can measure its KCBS value either by
reconstructing the whole density matrix of the state and then
calculating the projection value on each $|l_k\rangle$, or by
performing direct projection measurements on each $|l_k\rangle$
state. According to the projection measurement protocol for biphoton
qutrits suggested by Klyachko \textit{et al.} \cite{KCBS_2008}, one
should choose the $|l_k\rangle$ states to be neutrally polarized.
Thus, it is possible to write every biphoton state $|l_k\rangle$ as
a superposition of two orthogonally polarized single photons
$|s\rangle$ and $|t\rangle$
\begin{equation}\label{Neutrally_polarized}
|l_k\rangle\equiv|l_{st,k}\rangle=\frac{|s_k\rangle|t_k\rangle+|t_k\rangle|s_k\rangle}{\sqrt{2}}.
\end{equation}
With the usage of a Hanbury Brown-Twiss interferometer (i.e.,
separating the biphoton into two arms $\{1,2\}$ by a beam splitter
(BS)), one can project the biphoton onto the orthogonal separable
polarization state $|s\rangle_1|t\rangle_2$. Assuming that the
generated biphoton state occupies only the qutrit symmetric
subspace, and an ideal projection setup, the coincidence count rate
for the latter projection should be equal to the rate of projection
onto the $|t\rangle_1|s\rangle_2$ state, and their sum to the
biphoton projection rate onto the $|l_{st,k}\rangle$ state.

In practice, distinguishability between the two biphoton photons has
to be considered. As was shown by Adamson \textit{et al.}
\cite{Adamson_2007}, a realistic biphoton state is not perfect and
may contain some photon-pairs in the anti-symmetric singlet
subspace. Such unwanted photon pairs will be successfully projected
onto every pair of orthogonal polarizations $|s\rangle$ and
$|t\rangle$. As a result, the KCBS value calculated directly from
the projection measurements can even be higher than $\sqrt{5}$,
which is the upper limit allowed by quantum mechanics. Furthermore,
although the singlet subspace should be incoherent with the qutrit
subspace \cite{Adamson_2007}, imperfect symmetry between the two
arms of the projection setup may result in partially coherent terms
between the two subspaces. The appearance of these coherent terms
results in different coincidence rates for the
$|s\rangle_1|t\rangle_2$ and $|t\rangle_1|s\rangle_2$ projections,
and they deviate equally from bellow and above the rate that
corresponds to the ideal biphoton state. Therefore, with the direct
projection method, the coincidence rate $N_{s_k,t_k}$ should be
considered as the average of the $|s_k\rangle_1|t_k\rangle_2$ and
$|t_k\rangle_1|s_k\rangle_2$ rates in order to eliminate the
artificial coherence effect. Additionally, the relative part of the
generated singlet states should be subtracted from each coincidence
rate $N_{s_k,t_k}$. For that end, the singlet content should be
estimated using other projection measurements. Two possible ways can
be either by separating the biphoton to two different spatial modes
and measuring the visibility of a Hong-Ou-Mandel interference
\cite{Hong_1987}, or by performing a complete two-photon four
dimensional (4D) quantum state tomography (QST) procedure.

In order to measure the KCBS value of biphotons, we generated
$|1,1\rangle$ states and performed the projection protocol described
above. The experimental setup is shown in Fig. \ref{Fig1}a (see
elaboration in the supplementary material
\cite{Supplamentary_material}). A pulsed laser pumps two
perpendicular nonlinear crystals of equal length
\cite{Chekhova_2004}, and generates biphotons in the process of
spontaneous parametric down conversion. The generated biphotons are
in the state of $\
|\psi\rangle=\cos(2\delta)|2,0\rangle+\sin(2\delta)e^{i\varphi}|0,2\rangle,$
where $\delta$ is the angle of a half-wave plate (HWP) that controls
the pump beam polarization. Tilting a birefringent crystal that is
placed after the generating crystals for temporal compensation
controls the angle $\varphi$. Setting $\delta=22.5^\circ$ and
$\varphi=180^\circ$ results in a generation of a neutrally polarized
$|\psi_{pm}\rangle$ state. The $|\psi_{hv}\rangle$ ($|1,1\rangle$)
state is generated by passing the $|\psi_{pm}\rangle$ state through
another HWP oriented in an angle of $\alpha=22.5^\circ$.

\begin{figure}
\includegraphics[angle=0,width=86mm]{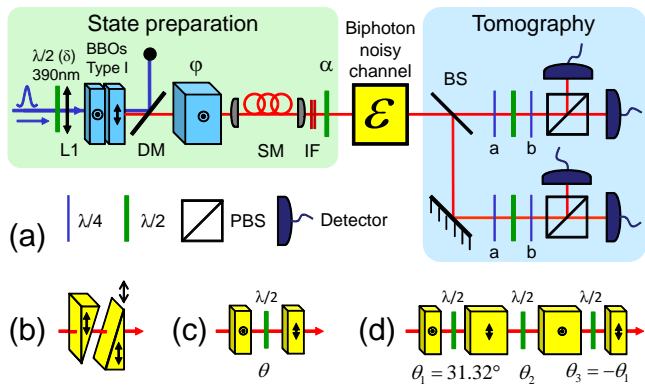}
\caption{\label{Fig1} (Color online) The experimental setup. (a)
Biphoton generation and characterization units: photon pairs are
generated in the BBO crystals, which are located after a lens (L1)
and a half-wave plate (HWP, $\lambda/2$) whose angle is $\delta$.
The down-converted photon-pairs pass through a dichroic mirror (DM),
a birefringent compensating crystal $(\varphi)$, a single-mode fiber
(SM), an interference bandpass filter (IF), and another HWP
($\alpha$). In the state characterization unit, the biphotons are
split probabilistically by a beam splitter (BS) into two ports. In
each port, the photons pass a sequence of a quarter-wave plate (QWP,
$\lambda/4$, a), a HWP, another QWP (b) and a polarizing beam
splitter (PBS) before being coupled into single-photon detectors.
The decoherence channels ($\mathcal{E}$) are plugged in only when
the noise affects on the biphotons are studied. (b) The one-field
(dephasing) channel, composed of two translatable quartz wedges. (c)
The two-field channel, composed of two perpendicularly oriented
identical 2\,mm thick calcite crystals. (d) The three-field
(isotropic) channel. This channel is composed of four crystals and
two fixed HWP. The thickness of the two outer (inner) crystals is
1\,mm (2\,mm). The fields' strength of both the two- and the
three-field channels is set by rotating the middle HWP.}
\end{figure}

In the characterization unit, the biphoton is probabilistically
split using a BS into two ports. Then, the two-port polarization
state is projected onto a separable two-qubit polarization state.
Thus we are able to perform the projection protocol, or
alternatively, to characterize the two-photon state using a complete
4D QST procedure \cite{James_2001}.

We generated a $|1,1\rangle$ biphoton state and reconstructed its
two-photon four-dimensional density matrix $\hat{\rho}$. We obtained
a fidelity of $93\pm1\%$ to the ideal state, where $99\pm1\%$ of the
state population occupied the symmetric subspace.

A KCBS value of $K=2.16\pm0.02>2$ is calculated from the measured
two-photon 4D density matrix $\hat{\rho}$, demonstrating violation
of $8\sigma$. The calculation was carried out by projecting the
symmetric subspace of $\hat{\rho}$ on a $|l_k\rangle$ quintuplet
(see elaboration on the $|l_k\rangle$ states in the supplementary
material \cite{Supplamentary_material}). Errors were calculated
using Monte-Carlo simulations assuming Poisson distribution for the
counts \cite{Kwiat_Tomo}. Performing the direct projection
measurements, we measured a violation of $4\sigma$, with a KCBS
value of $K=2.17\pm0.04$. The singlet population which is required
for the direct projection protocol was taken from the measured 4D
density matrix of the biphoton. Here, the error value was estimated
assuming Poisson distribution for the projections onto the
$|l_k\rangle$ states, and by performing Monte-Carlo simulations to
derive the error for the singlet subspace population.

In order to compare the resilience of two fundamental characters of
quantum mechanics, contextuality and nonlocality, we explore the
effects of decoherence on the ability of biphotons to exhibit these
nonclassical effects. Neutrally polarized biphotons were transmitted
through different types of quantum noisy channels and the resulting
two-photon 4D density matrices were reconstructed. Specifically, we
compered between the final states ability to post-selectively
violate the CHSH nonlocality inequality and its ability to
demonstrate contextuality by violating the KCBS inequality
(\ref{KCBS_eq}). The CHSH operator $S_{CHSH}$ is a function of
$\hat{\rho}$ obtained by four different projection measurements, and
nonlocality is demonstrated when the inequality
$|\langle{S}_{CHSH}\rangle|\leq2$ is violated \cite{CHSH_1969}. For
a measured output state $\hat{\rho}$, the maximal value of
$|\langle{S}_{CHSH}\rangle|$ was calculated
 using the Horodecki criterion \cite{Horodecki_1995}, and the maximal KCBS
value was obtained using a numerical search for the $|l_k\rangle$
quintuplet that gives the highest KCBS value.

\begin{figure*}[tbp]
\includegraphics[angle=0,width=\textwidth]{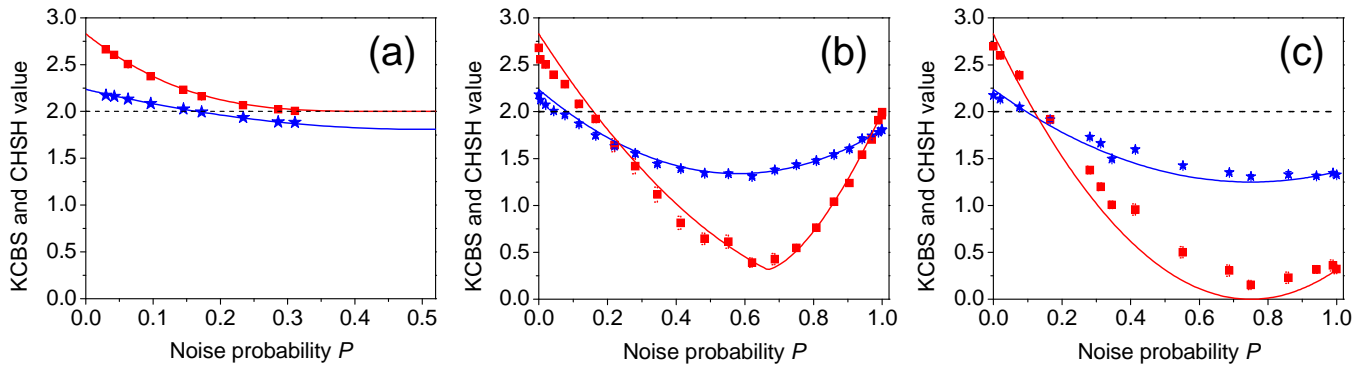}
\caption{\label{Fig2} (Color online) The KCBS and the CHSH values in
the presence of unital noise. The maximal possible expectation
values for the KCBS (blue pentagrams) and for the CHSH (red squares)
operators of the output biphoton states are presented as a function
of the noise probability. Data is presented for the three principal
unital channels: (a) dephasing, (b) two-field and (c) isotropic
channels. Solid lines represent the theoretical predictions. Dashed
lines represent the value 2 - the upper bound allowed by any
classical theory for the expectation values of both operators.}
\end{figure*}

The operation $\mathcal{E}$ of a quantum channel on a state
$\hat{\rho}$ is usually defined by a map
$\hat{\rho}'=\mathcal{E}(\hat{\rho})$, where a unital channel
satisfies $\mathcal{E}(\hat{I})=\hat{I}$. We implemented three
principal types of single-photon unital channels, with a controlled
amount of noise, operating independently on both photons. The first
channel is a one-field dephasing channel, where its single-qubit
operation is described by
\begin{equation}\label{dephasing_channel}
\mathcal{E}(\hat{\rho})=(1-P)\hat{\rho}+P\sigma_1\hat{\rho}\sigma_{1}\,.
\end{equation}
$P$ is the probability to apply the channel, and $\sigma_1$,
$\sigma_2$, and $\sigma_3$ are the Pauli matrices. The channel is
composed of two translatable quartz wedges (see Fig. \ref{Fig1}b)
that create a variable temporal delay between the polarization modes
\cite{Branning_2000}. The second channel is a two-field channel
\begin{equation}\label{two_field_channel}
\mathcal{E}(\hat{\rho})=(1-P)\hat{\rho}+\frac{P}{2}\sigma_1\hat{\rho}\sigma_{1}+\frac{P}{2}\sigma_2\hat{\rho}\sigma_{2},
\end{equation}
and the third one is an isotropic three-field channel
\begin{equation}\label{isotropic_channel}
\mathcal{E}(\hat{\rho})=(1-P)\hat{\rho}+\frac{P}{3}\sigma_1\hat{\rho}\sigma_{1}+\frac{P}{3}\sigma_2\hat{\rho}\sigma_{2}+\frac{P}{3}\sigma_3\hat{\rho}\sigma_{3}.
\end{equation}
The second and third channels are composed of a sequence of fixed
birefringent calcite crystals and HWPs \cite{Shaham,Shaham_iso_depo}
(see Figs. \ref{Fig1}c and \ref{Fig1}d, respectively). Each crystal
entangles the polarization modes of a photon with its internal
temporal degrees of freedom (DOFs). Decoherence occurs if the photon
detection is insensitive to the temporal delays, practically
averaging over these DOFs \cite{Kwiat_Dephasing,Shaham}. For both
channel types, control over the noise probability $P$ is achieved by
rotating the corresponding HWPs to different angle settings
\cite{Shaham} (for further elaboration see supplementary material
\cite{Supplamentary_material}).

The KCBS and the CHSH values of the output biphoton state as a
function of the noise parameter $P$ of the different channels are
shown in Fig. \ref{Fig2}. Solid lines represent the theoretical
predictions for ideal $|\psi\rangle$ initial states. For the
dephasing channel, the noise probability $P$ was deduced from the
purity of the measured state (details in the methods section). For
the two other channels, the noise probability $P$ was derived from
the corresponding HWP settings. Thus the theoretical predictions for
the two-field and the isotropic channels are without any fit
parameters.

As can be seen by comparing the state dynamics of the different
channels presented in Fig. \ref{Fig2}, by adding more fields to the
decoherence process, the KCBS and CHSH values can reach lower
values. It is evident that for all three channels, the ability to
violate the KCBS inequality is lost earlier than the ability to
exhibit nonlocality. Specifically, for the dephasing channel (Fig.
\ref{Fig2}a), nonlocality is preserved for all noise probabilities
(except for the extreme case of $P=0.5$), while the ability to
exhibit KCBS correlations vanishes at $P\sim0.17$. The results also
demonstrate quantum hierarchy; a two-photon state that can exhibit
KCBS correlations can also exhibit nonlocality, while the inverse is
not true \cite{Soeda_2013}. The deviation between measurement and
theory for the isotropic channel (Fig. \ref{Fig2}c) is explained by
the usage of narrower interference bandpass filters (3\,nm instead
of 5\,nm) that on one hand increased the fidelity of the initial
state, but on the other, extended its coherence time. Since the
noise channel operation relies on the finite coherence time of the
incoming light, a longer coherence time reduces its depolarization
strength. Nevertheless, The hierarchy described above holds also for
the experimental results of this channel, as predicted.

In conclusion, we studied the ability of a symmetric two-photon
state to exhibit quantumness. In the first part of the work, the
KCBS value of a biphoton qutrit state was measured. The direct
projection method was extended to the case where some of the
biphoton population is distinguishable, and thus it resides also in
the singlet subspace. The measured KCBS value is above the limit
allowed by any classical theory. Assuming fair sampling and that the
projections indeed represent compatible measurements
\cite{Guhne_2010}, it has been shown that there exists no joint
probability distribution that can explain the measurement outcomes
of an experiment performed on the investigated qutrit state. In the
second part of this work, we transmitted biphotons through three
principal types of unital channels, and observed the degradation in
the state ability to violate the KCBS and the CHSH inequalities as
the noise level is increased. We demonstrated the predicted
hierarchy between the two properties, i.e., KCBS contextuality
implies nonlocality.

From a quantum information point of view, nonlocality is an
essential resource for quantum computation tasks. Recently, there
has been a growing interest in the study of quantum contextuality as
a resource for quantum information applications
\cite{Horodecki_2010,Howard_2014,Gedik_2014,Dogra_2014}. Our results
show that when the two photons of a biphoton experience a unital
process, their nonlocal correlations are more robust than their KCBS
contextual correlations. This leads to two possible future research
directions. It is interesting to search for contextual based-on
quantum protocols, that use different correlations from the KCBS
correlations. Can these correlations be more robust against noise,
even with respect to nonlocal based-on quantum protocols?
Additionally, one can wonder if the observed fragility of the KCBS
correlations hints that KCBS-based quantum technologies will exhibit
superior performance to current entanglement-based approaches.

\begin{acknowledgments}
We thank the Israeli Ministry of Science and Technology for
financial support and the Israeli Science Foundation for supporting
this work under Grants 546/10 and 793/13.
\end{acknowledgments}

\clearpage

\begin{center}
\textbf{\Large{Supplementary Material}}
\end{center}

%


\section*{Experimental setup}

Biphotons are collinearly generated in the process of spontaneous
parametric down conversion. Using a lens of 30\,cm focal length
(L1), a pulsed 390\,nm pump laser was focused onto two
perpendicularly oriented 1\,mm thick type-I
$\beta-\textrm{BaB}_{2}\textrm{O}_{4}$ (BBO) crystals. After the
crystals, the down-converted signal is separated from the pump beam
using a dichroic mirror (DM). A half-wave plate (HWP) denoted by
($\delta$) is placed before the generating crystals in order to
control the ratio between the pumping power of each crystal. Thus,
the generated neutrally polarized state is
$|\psi\rangle=(|2,0\rangle+e^{i\varphi}|0,2\rangle)/\sqrt{2}$. By
tilting the temporal walk-off compensating crystal, which is placed
after the generating crystals, the $\varphi$ angle is controlled.
The state is filtered spatially using a single-mode fiber (SM), and
spectrally by 3\,nm or 5\,nm interference bandpass filters (IF). The
HWP denoted by $\alpha$ can rotate a $|\psi_{pm}\rangle$ state to
the $|1,1\rangle$ state if $\alpha=22.5^\circ$. In the state
tomography unit, the photons are split probabilistically at a beam
splitter (BS). The required settings for the projection protocol and
for the quantum state tomography procedure are achieved by a
sequence of a quarter-wave-plate (QWP), a HWP, another QWP and a
polarizing beam splitter (PBS), that are placed before the
single-photon detectors of each port.

The quantum noise channels were placed before the BS. The noise
probability of the dephasing channel is controlled by the
translation of two quartz wedges in different directions, in order
to change the optical length inside the birefringent wedges. The
time delay range that can be set between the two polarizations is
$\sim380\,fs$. The measurements of Fig. 2a in the main text were
taken through all this range. The noise probability $P$ was deduced
from the purity ($\Pi=\textrm{Tr}(\hat{\rho}^2)$) of the measured
state. Assuming that the initial state is ideal, that the channel is
a dephasing channel, and that $P\leq0.5$, we use the relation
$P=(1-(2\Pi-1)^{\frac{1}{4}})/2$ to obtain $P$. The noise
probabilities of the two-field and the isotropic channels are
controlled by the rotation of the corresponding channel middle
wave-plate. For both channels $P=\textrm{sin}^2(2\theta)$, where for
the isotropic channel $\theta$ is $\theta_2$. The initial state for
the one- and three-field channels was $|\psi_{pm}\rangle$. Due to
technical considerations, a $|\psi_{rl}\rangle$ state was used for
the two-field channel.

\section*{Projection settings for the KCBS measurements}

\begin{figure}[b]
\includegraphics[angle=0,width=86mm]{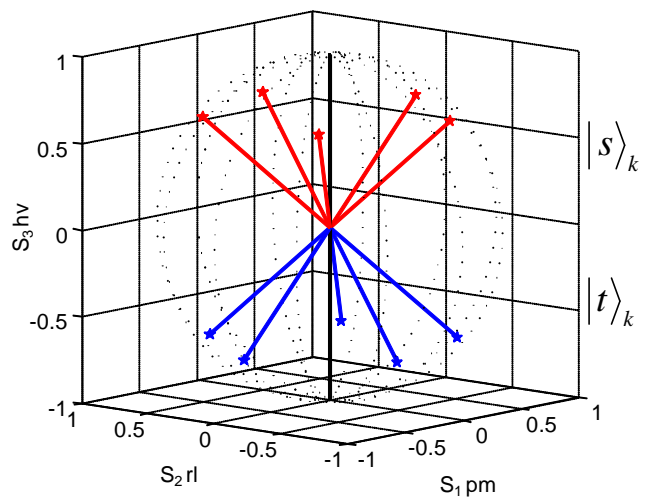}
\caption{\label{Fig3} (Color online) The polarization states used
for the direct projection protocol. The 5 $|s_k\rangle$ states,
along with their antipodes $|t_k\rangle$ states are presented in the
Stokes representation. All 10 states have the same distance from the
black solid line that connects the horizontal (h) and the vertical
(v) polarizations, located at the poles of the Poincar\'{e} sphere.}
\end{figure}

The five $|l_{st,k}\rangle$ states used in the projection protocol
have the coefficients
$\{\alpha_0,\alpha_1,\alpha_2\}_k=\{\sin(\theta)/\sqrt{2},\cos(\theta)e^{i\varphi_k},-\sin(\theta)e^{2i\varphi_k}/\sqrt{2}\}$,
where $\theta=\cos^{-1}(\frac{1}{\sqrt[4]{5}})\simeq48.03^\circ$ for
every $|l_{st,k}\rangle$ and $\varphi_k=\frac{4}{5}\pi k$. These
states are pairwise orthogonal
$|l_{st,k}\rangle\perp|l_{st,k+1}\rangle$ ($k+1$ modulo 5) and
satisfy the maximal violation of Eq. (2) of the main text for an
initial $|1,1\rangle$ state. The corresponding  orthogonal single
photon polarizations $|s_k\rangle$ and $|t_k\rangle$ that compose
every $|l_{st,k}\rangle$ state are presented in Fig. \ref{Fig3}.
Note that these 10 polarization states form two pentagrams with a
symmetry axis that connects the $|h\rangle$ and $|v\rangle$
antipodes, which compose the $|1,1\rangle$ state. In order to
project the $|1,1\rangle$ state onto all the
$|s_k\rangle_1|t_k\rangle_2$ and $|t_k\rangle_1|s_k\rangle_2$
states, as required by the projection protocol, the three wave
plates in both ports after the BS were tuned to the angles listed in
table \ref{table1}. Then, the coincidence counts of
$|h\rangle_1|v\rangle_2$ ($|v\rangle_1|h\rangle_2$) detectors were
measured, and the projection results of $|s_k\rangle_1|t_k\rangle_2$
($|t_k\rangle_1|s_k\rangle_2$) were deduced.

\begin{table}[h]
  \centering
  \begin{tabular} {| l | r | r | r | r | r |}
  \hline
  \multicolumn{1}{|c|}k & \multicolumn{1}{|c|}1 & \multicolumn{1}{|c|}2 & \multicolumn{1}{|c|}3 & \multicolumn{1}{|c|}4 & \multicolumn{1}{|c|}5 \\ \hline \hline
  $\textrm{QWP (a)}$  & $3.92^\circ $ & $-9.91^\circ$ & $12.01^\circ$ & $-9.91^\circ$ & $3.92^\circ$ \\ \hline
  $\textrm{HWP}$    & $-11.39^\circ$ & $+6.92^\circ$ & \multicolumn{1}{|c|}{$0$} & $-6.92^\circ$ & $11.39^\circ$ \\ \hline
  $\textrm{QWP (b)}$  & $-3.92^\circ$ & $9.91^\circ$ & $-12.01^\circ$ & $9.91^\circ$ & $-3.92^\circ$ \\ \hline

  \end{tabular}
  \caption{The wave-plate angles for the direct projection measurements of the KCBS value.
  In order to successfully project a biphoton on a certain $|s_k\rangle_1|t_k\rangle_2$ or $|t_k\rangle_1|s_k\rangle_2$ state,
  each photon coming from the two ports of the BS has to pass
  through a QWP (a), a HWP, and another QWP
  (b), whose angles are shown in this table for every $k$. Then,
  the photons pass through a polarizer in each arm, oriented horizontally or
  vertically, where the orientation of the two polarizers is
  perpendicular, i.e., either $|h\rangle_1|v\rangle_2$ or $|v\rangle_1|h\rangle_2$.}
  \label{table1}
\end{table}


\end{document}